\documentclass[fleqn,twoside]{article}
\usepackage{amsmath}
\usepackage{espcrc2}

\newcommand{\rd}[1]{\mathop{\mathrm{d}#1}}
\newcommand{\fract}[2]{{\textstyle\frac{#1}{#2}}}
\newcommand{\grad}{\vec\nabla}

\newcommand{\nc}{noncommuting}

\newcommand{\vA}{\vec A}

\newcommand{\bra}[1]{\bigl\langle #1 \bigr| }
\newcommand{\ket}[1]{\bigl| #1 \bigr\rangle}
\newcommand{\braket}[2]{\bigl\langle #1 \big|  #2 \bigr\rangle}

\newcommand{\numeq}[2]{\begin{equation}
#2
\label{#1}
\end{equation}}
\newcommand{\refeq}[1]{(\ref{#1})}

\let\vec\boldsymbol
\let\eps\varepsilon
\let\epsilon\varepsilon
\let\phi\varphi
\let\hat\widehat

\title{Noncommuting fields and non-Abelian fluids}

\author{R. Jackiw\address[CTP]{Center for Theoretical Physics\\ 
Massachusetts Institute of Technology\\ 
Cambridge, MA 02139-4307}%
        \thanks{This work is supported in part by funds provided by the US
Department of Energy (DOE) under cooperative research agreement
\#DF-FC02-94ER40818}}
       
\begin{document}

\begin{abstract}
The original ideas about noncommuting coordinates are recalled. The connection between U(1)
gauge fields defined on noncommuting coordinates and fluid mechanics is explained.
Non-Abelian fluid mechanics is described.
\vspace{1pc}
\end{abstract}

\maketitle

\noindent {\bfseries\large I}\enspace 
The idea that configuration-space coordinates may not commute,
\numeq{e1}{
[x^i, x^j] = i\theta^{ij}
}
where $\theta^{ij}$ is a constant, antisymmetric two-index object, has arisen recently from
string theory, but in fact it has an older history.  Like many interesting quantum-mechanical
ideas, it was first suggested  by Heisenberg, in the late 1930s, who reasoned that coordinate
noncommutativity would entail a coordinate uncertainty and could ameliorate short-distance
singularities, which beset quantum fields. He told  his idea to Peierls, who eventually made use
of it when analyzing electronic systems in an external magnetic field, so strong that projection
to the lowest Landau level is justified.   But this phenomenological realization of
Heisenberg's idea, which I shall discuss presently, did not address  issues in fundamental
science, so Peierls told Pauli about it, who in turn told Oppenheimer, who  asked his student
Snyder to work it out and this  led to the first published paper on the subject~\cite{r1}. 

The coordinate noncommutativity in the lowest Landau level is very similar to today's
string-theory origins of noncommutativity -- both rely on the presence of a strong
background field. Also, thus far, it is the only physically realized example of noncommuting
coordinates, so let me describe it in a little detail~\cite{r2}.
We consider the motion of a charged ($e$) and massive ($m$) particle in a constant magnetic
field ($B$) pointing along the $z$~direction.  All interesting physics is in the $x$--$y$ plane. The
Lagrangian for this planar motion is
\numeq{n2}{
L = \fract12 m(\dot x^2 +\dot y^2) +   e  (\dot x A_x + \dot y A_y) - V(x,y) 
}
where the vector potential $\vA$ can be chosen as $(0, xB)$ and $V(x,y)$ describes additional
interactions (``impurities''). In the absence of $V$, the quantum spectrum consists of the
well-known Landau levels $\ket{n,d}$, where $n$ indexes the level's energy eigenvalue,
and~$d$ describes the infinite degeneracy of each level. The separation between levels is
$O(B/m)$, so that in the strong magnetic field limit only the lowest Landau level $\ket{0,d}$ is
relevant. But observe that the large $B$ limit corresponds to small~$m$, so projection on the
lowest Landau level is also achieved by setting $m$ to zero in \refeq{n2}. In that limit the
Lagrangian \refeq{n2}, in the chosen gauge, becomes
\numeq{n3}{
L_{\ell L\ell} =  e  Bx\dot y - V(x,y)\ . 
}
This is of the form $p\dot q - H(p,q)$, and immediately identifies $e Bx$ and $y$ as
canonical conjugates, leading in the usual way to the commutator
\numeq{n4}{
[x,y] = - i \frac{\hbar}{eB}\ .
}
(The ``Peierls substitution'' consists of determining the effect of the impurity by computing the
eigenvalues of $V(x,y)$, where $x$ and $y$ are noncommuting.
There is much more to be said about this. For example, the issue of ordering the (now) \nc\
arguments $(x,y)$ of $V$ must be settled. Also one needs to understand the behavior of wave
functions. Before the (phase-space reductive) strong-$B$ limit they depend on \emph{both}
$x$ and~$y$, while after the limit they can depend only on \emph{one} of the two \nc\
variables, all the time retaining  their normalization. All this, as well as other matters, is
explained in Ref.~\cite{r2}.)

For another perspective, consider calculating the lowest Landau level  matrix elements of
the $[x,y]$ commutator 
\numeq{n5}{
\bra{0,d} xy - yx \ket{0,d'} = M(d,d') - M^* (d',d)
}
where
\numeq{n6}{
 M(d,d') = \bra{0,d} xy \ket{0,d'} \ .
}
We evaluate \refeq{n6} by inserting intermediate states in product $xy$:
\numeq{n7}{
 M(d,d') = \sum_s \bra{0,d} x  \ket{s} \bra{s}  y \ket{0,d'}\ . 
}
If the sum is over \emph{all} the degenerate Landau levels, then one finds that \refeq{n5}
vanishes: $x$ and $y$ do commute! But if one pretends that the world is restricted to the
lowest Landau level and includes only that level (with its degeneracy) in the intermediate
state sum 
\numeq{n8}{
M_{\ell L\ell} = \sum_{d''} \bra{0,d} x  \ket{0,d''} \bra{0,d''}  y \ket{0,d'}
}
one finds that on this truncated state space, eq.~\refeq{n5} becomes consistent with \refeq{n4}:
\numeq{n9}{
\bra{0,d} [x,y] \ket{0,d'} = -i \frac{\hbar}{eB} \braket{0,d}{0,d'} \ . 
}

A generalization now suggests itself: Suppose one retains not just the lowest Landau level, but
includes the first~$N$ levels (setting $N=0$ for the lowest level)~\cite{r3}. In this case one
should consider all levels $n\leq N$, both as external states and contributing to the
intermediate state sum in \refeq{n7}. The result is that the matrix elements of the coordinate
commutator vanish unless the external levels ($n,n'$) coincide, $n=n'$, and for the coincident
case the only nonvanishing matrix element is at the highest Landau level $n=n'=N$:
\begin{gather} \label{n9a}
\bra{N,d} [x,y] \ket{N,d'}\\
\qquad\qquad\qquad  = -i \frac{\hbar}{eB}(N+1) \braket{N,d}{N,d'} \ . \nonumber
\end{gather}
As $N\to\infty$, more and more levels are included, the highest level decouples and the
coordinates commute.

\bigskip \noindent {\large\bfseries  II}\enspace  
Let me now return to the general and abstract problem of
noncommuting coordinates.  When confronting the noncommutativity postulate~\refeq{e1}, it
is natural to ask which (infinitesimal) coordinate transformations 
\numeq{e2}{
\delta x^i = -f^i (x) 
}
leave \refeq{e1} unchanged. 
The answer is that the (infinitesimal) transformation vector function $f^i(x)$ must be
determined by a scalar $f(x)$ through the expression~\cite{r4}
\numeq{e3}{
f^i (x) = \theta^{ij} \partial_j f(x) \ . 
}
Since then  $\partial_i f^i(x) = 0$,   these are recognized as volume-preserving
transformations. [They do not exhaust all volume-preserving transformations, except in two
dimensions. In dimensions greater  two,
\refeq{e3} defines a subgroup of volume-preserving transforms that also leave $\theta^{ij}$
invariant.]

The volume-preserving transformations form the link between noncommuting coordinates
and fluid mechanics. Since the theory of fluid mechanics is not widely known outside the circle
of fluid mechanicians, let me put down some relevant facts \cite{r5}.  There are two, physically
equivalent descriptions of fluid motion: One is the Lagrange formulation, wherein the fluid
elements are labeled, first by a discrete index~$n$: $\vec X_n(t)$ is the position as a function of
time of the $n$th fluid element.  Then one passes to a continuous labeling variable $n \to \vec
x: \vec X_n(t)\to \vec X (t,\vec  x)$, and 
$\vec x$ may be taken to be the position of the fluid element at initial time $\vec X(0,\vec x) =
\vec x$. This is a comoving description. Because labels can be arbitrarily rearranged, without
affecting physical content, the continuum description is invariant against volume-preserving
transformations of $\vec x$, and in particular, it is invariant against the specific
volume-preserving transformations
\refeq{e3}, provided the fluid coordinate $\vec X$ transforms as a scalar:
\numeq{e4}{
\delta_f  \vec X  = f^i (\vec x) \frac\partial{\partial x^i} \vec X = \theta^{ij} \partial_i \vec
X\partial_j f\ . 
 }

The common invariance of Lagrange fluids and of noncommuting coordinates is a strong hint of
a connection between the two.

Formula \refeq{e4}  takes  a very suggestive form when we rewrite it in terms of a bracket
defined for functions of $\vec x$ by 
\numeq{e5}{
\bigl\{ \mathcal O_1(\vec x), \mathcal O_2(\vec x)\bigr\} =\theta^{ij} \partial_i\mathcal
O_1(\vec x)
\partial_j\mathcal O_2(\vec x)\ .
} 
Note that with this bracket we have
\numeq{e6}{
\bigl\{x^i,  x^j\bigr\} = \theta^{ij}\ .
} 
So we can think of bracket relations as classical precursors of commutators for a
noncommutative field theory -- the latter obtained from the former by replacing brackets by
$-i$~times commutators, \`a la Dirac. More specifically, we shall see that the noncommuting field
theory that emerges from the Lagrange fluid is a noncommuting U(1) gauge theory. 

This happens when the following steps are taken. We define the evolving portion of $\vec X$ by 
\numeq{e7}{
X^i (t,\vec x) = x^i + \theta^{ij} \hat A_j (t,\vec x)\ .
}
(It is assumed that $\theta^{ij}$ has an inverse.)
Then \refeq{e4} is equivalent to the suggestive expression
\numeq{e8}{
\delta_f \hat A_i = \partial_i f + \bigl\{\hat A_i, f\bigr\}\ .
}
When the bracket is replaced by $(-i)$ times the commutator, this is precisely the gauge
transformation for a noncommuting U(1) gauge potential $\hat A_i$. Moreover, the gauge field
$\hat F_{ij}$ emerges from the bracket of two Lagrange coordinates
\begin{gather}
\bigl\{ X^i, X^j\bigr\} = \theta^{ij} + \theta^{im} \theta^{jn} \hat F_{mn} \label{e9}\\
\hat F_{mn}  = \partial_m \hat A_n - \partial_n \hat A_m + \bigl\{\hat A_m, \hat  A_n\bigr\}\ . 
\label{e10}
\end{gather}
Again \refeq{e10} is recognized from the analogous formula
in noncommuting gauge theory.

\bigskip \noindent {\large\bfseries  III}\enspace  
What can one learn from the parallel formalism for a Lagrange
fluid and a noncommuting gauge field? One result that has been obtained addresses the
question of what is  a gauge field's covariant response to a coordinate transformation. This
question can be put already for commuting, non-Abelian gauge fields, where conventionally
the response is given in terms of a Lie derivative $L_f$:
\begin{gather}
\delta_f x^\mu = - f^\mu(x) \label{e11}\\
\delta_f A_\mu = L_f A_\mu \equiv f^\alpha \partial_\alpha A_\mu + \partial_\mu f^\alpha
A_\alpha
 \ . 
\label{e12}
\end{gather}
But this implies
\begin{gather} \label{e13}
\delta_f F_{\mu\nu} = L_f F_{\mu\nu} \\
\qquad\quad \equiv f^\alpha \partial_\alpha F_{\mu\nu} + 
\partial_\mu f^\alpha F_{\alpha\nu} +  \partial_\nu f^\alpha F_{\mu\alpha}
\nonumber
\end{gather}
which is not covariant since the derivative in the first term on the right is not the covariant
one. Thus we consider the conventional approach to be defective because it does not preserve
gauge invariance. The cure in this, commuting, situation has been given some time
ago~\cite{r6}:  Observe that
\refeq{e12} may be equivalently presented as 
\begin{align}
\delta_f A_\mu &= L_f A_\mu = f^\alpha \bigl(
\partial_\alpha A_\mu -\partial_\mu A_\alpha - i [A_\alpha, A_\mu]
\bigr) \nonumber\\
&\qquad{}+ f^\alpha \partial_\mu A_\alpha  - i [  A_\mu, f^\alpha A_\alpha] + 
\partial_\mu f^\alpha A_\alpha\nonumber\\
 &= f^\alpha F_{\alpha\mu} +  D_\mu (f^\alpha A_\alpha)\ . \label{e14}
\end{align}
Thus, if the coordinate transformation  generated by $f^\alpha$ is supplemented by a  gauge
transformation generated by $-f^\alpha A_\alpha$, the result is a gauge covariant coordinate
transformation
\numeq{e15}{
\delta'_f A_\mu = f^\alpha F_{\alpha\mu}
}
and the modified response of $F_{\mu\nu}$ involves the gauge-covariant Lie derivative $L'_f$:
\begin{gather}
\delta'_f F_{\mu\nu} = L'_f F_{\mu\nu}\label{e16}\\
\qquad\quad \equiv f^\alpha D_\alpha F_{\mu\nu} + 
\partial_\mu  f^\alpha F_{\alpha\nu} +  \partial_\nu f^\alpha  F_{\mu\alpha} \ .
\nonumber
\end{gather}

In the noncommuting situation, loss of covariance in the ordinary Lie derivative is even
greater, because in general the coordinate transformation functions $f^\alpha$ do not
commute with the fields $\hat A_\mu, \hat F_{\mu\nu}$; moreover, multiplication of
$x$-dependent quantities is not a covariant operation. All these issues can be addressed and
resolved by considering them in the fluid mechanical context, at least, for linear and
volume-preserving diffeomorphisms. The analysis is technical and I refer you to the published
papers~\cite{r4,r7}. The final result for the covariant coordinate transformation on the
noncommuting gauge potential $\hat A_\mu$, generated by $f^\alpha (x)$, is 
\begin{gather}
\delta'_f \hat A_\mu = \fract12 \bigl\{
f^\alpha(X), \hat F_{\alpha\mu}\bigr\}_+ \label{en25}\\
\qquad\qquad\qquad\qquad   \mbox{ plus reordering terms.} \nonumber
\end{gather}
Note that the generating function $f^\alpha(X)$ enters the
anticommutator $\{\ ,\ \}_+$ with covariant argument~$X$. $f^\alpha$ is restricted to
be either linear or volume-preserving; in the latter case there are reordering
terms, whose form is explicitly determined by the fluid mechanical
antecedent~\cite{r4,r7}.

\bigskip \noindent {\large\bfseries  IV}\enspace 
 Next, I shall discuss the Seiberg-Witten map~\cite{r8}, which can be
made very transparent by the fluid analogy.  The Seiberg-Witten map replaces the
noncommuting vector potential $\hat A_\mu$  by a nonlocal function of a commuting potential
$A_\mu$ and of~$\theta$;  i.e., the former is viewed as a function of the latter.  The relationship
between the two follows from the requirement of stability against gauge transformations: a
noncommuting gauge transformation  of the noncommuting gauge potential should be
equivalent to a commuting gauge transformation on the commuting vector potential on which
the noncommuting potential depends. Formally:
\numeq{en26}{
\hat A_\mu(A + \rd \lambda) = \hat A_\mu^{G(a,\lambda)} (A)\ .
}
Here $\lambda$ is the Abelian gauge transformation function that transforms the Abelian,
commuting gauge potential $A_\mu$; $G(A,\lambda)$ is the noncommuting gauge function that
transforms the noncommuting gauge potential $\hat A_\mu$. $G$ depends on $A_\mu$
and~$\lambda$, and one can show that it is a noncommuting 1-cocycle~\cite{r9}.

Moreover, when the action and the equations of motion of
the noncommuting theory are transformed into commuting variables, the dynamical content
is preserved: the physics described by noncommuting variables is equivalently described by
the commuting variables, albeit in a complicated, nonlocal fashion. 

  The Seiberg-Witten map is intrinsically interesting in the unexpected equivalence that it
establishes. Moreover, it is practically useful for the following reason. It is difficult to extract
gauge-invariant content from a noncommuting gauge theory because quantities constructed
locally from $\hat F_{\mu\nu}$ are not gauge invariant. To achieve gauge invariance, one must
integrate over space-time. Yet for physical analysis one wants local quantities: profiles of
propagating waves, etc.  Such local quantities can be extracted in a gauge-invariant manner
from the physically equivalent, Seiberg-\linebreak[1]Witten--mapped
commutative gauge theory.

For example, colleagues and I addressed the following question~\cite{r10}. Consider free,
commuting Maxwell theory, and its \nc\ counterpart. The physical phenomena implied by the
former are well known: they are the electromagnetic waves, e.g., the plane waves. Can one
find analogous phenomena in the \nc\ theory, compare the two, and thereby set some
experimental limits on the noncommutativity?
 
The previously mentioned difficulty is now evident: Waves are local excitations, but no
meaning can be attached to local quantities in the \nc\ theory because they are gauge variant.
On the other hand, gauge-invariant \nc\ quantities involve integration over space-time, but we
would lose detailed physical information about Maxwell waves if their profiles are integrated. 


Here the Seiberg-Witten map suggests a solution. Working to $O(\theta)$,  we express the \nc\
fields in terms of commuting ones: 
\begin{subequations}
\begin{align}
\hat A_\mu &= A_\mu - \fract12 \theta^{\alpha\beta} A_\alpha (\partial_\beta A_\mu -
F_{\beta\mu}) \label{e28a}\\
\hat F_{\mu\nu} &= F_{\mu\nu} + \theta^{\alpha\beta} F_{\alpha\mu} F_{\beta\nu} -
\theta^{\alpha\beta} A_\alpha \partial_\beta F_{\mu\nu} \label{e28b}
\end{align}
\end{subequations}
where $F_{\mu\nu}$ is the Abelian field strength, the curl of $A_\mu$.
These profiles satisfy the condition \refeq{en26} with
\numeq{e29n}{
G(A,\lambda) = \lambda - \theta^{\alpha \beta} A_\alpha \partial_\beta \lambda\ . 
}
Also \refeq{e28a} solves to $O(\theta)$ a differential equation that can be derived from
\refeq{en26}:
\begin{align}
\frac{\partial \hat A_\mu}{\partial \theta^{\alpha\beta}} &=
  -\fract18 \{\hat A_\alpha, \partial_\beta \hat A_\mu - \hat F_{\beta\mu}\}_+
\label{e30n}\\
&\qquad{}+ \fract18 \{\hat A_\beta, \partial_\alpha \hat A_\mu - \hat
F_{\alpha\mu}\}_+\ .\nonumber
\end{align}
Note however that when
$\theta^{\alpha\beta}$ is larger than $(2\times2)$, \refeq{e30n} entails a \emph{set} of
equations, but they are not integrable. This means that when one constructs $\hat A_\mu
(\theta\neq0)$ from $\hat A_\mu (\theta=0) = A_\mu$, the construction depends on the path
in $\theta$-space. We shall return to this point. 

The commutative action, which emerges when \refeq{e28b} is inserted in the \nc\ Maxwell
action,   reads~\cite{r11}
\begin{gather}
I  =  \int \rd{^4x} \Bigl(-\fract{1}{4} F^{\mu\nu}F_{\mu\nu} \label{e31n}\\
\qquad\quad{}
+\fract{1}{8}\theta^{\alpha\beta}F_{\alpha\beta}F_{\mu\nu}F^{\mu\nu}
-\fract{1}{2}\theta^{\alpha\beta}F_{\mu\alpha}F_{\nu\beta}F^{\mu\nu}\Bigr)\ .
\nonumber
\end{gather}
The ``Maxwell'' equations are
\begin{subequations}\label{e32n}
\begin{align}
\frac{\partial}{\partial t} \vec B + 
\grad \times \vec E &= 0 \label{e32a} \\
\grad \cdot \vec B &=0\ . \label{e32b}
\end{align}
\end{subequations}
These are not dynamical; they merely express the fact that electric and magnetic fields ($E^i=
F^{i0}$,
$B^i = -\fract12 \eps^{ijk} F_{jk}$) are given in terms of potentials. The  remaining
equations, which follow from \refeq{e31n}, can be given a Maxwell form
\begin{subequations}\label{e33n}
\begin{align}
\frac{\partial}{\partial t} \vec D - 
\grad \times \vec H &= 0 \label{e33a} \\
\grad \cdot \vec D &=0\ . \label{e33b}
\end{align}
\end{subequations}
with constituent relations following from \refeq{e31n}:
\begin{subequations}\label{e34n}
\begin{align}
\vec D &= (1 - \vec\theta\cdot\vec B ) \vec E + 
(\vec\theta\cdot \vec E) \vec B + (\vec E \cdot 
\vec B) \vec\theta \label{e34a} \\
\vec H &= (1 - \vec\theta\cdot\vec B) \vec B\nonumber\\
&\qquad\qquad{} + 
\fract12(\vec E^2 - \vec B^2) \vec\theta - 
(\vec\theta\cdot \vec E) \vec E\ .
\label{e34b}
\end{align}
\end{subequations}
Here we have assumed that $\theta^{\alpha\beta}$ has no temporal component:
$\theta^{0\alpha}=0$, $\theta^{ij} = \eps^{ijk}\theta^k$. 

It turns out that [to $O(\theta)$] the nonlinear equations \refeq{e32n}--\refeq{e34n} still
possess plane wave solutions. However, in the presence of a background magnetic field $\vec
b$ the dispersion law between frequency~$\omega$ and wave number $\vec k$ is modified
from the Maxwell result
$\omega= k$; we find
\numeq{e35n}{
\omega = k (1-\vec \theta_T \cdot \vec b_T) 
}
where the subscript ``$T$'' denotes components transverse to the direction of
propagation~$\vec k$. 

Evidently, Lorentz symmetry is violated: the velocity of light has changed, increasing or
decreasing over the Maxwell value. In principle this should lead to a fringe shift in a
Michaelson-Morley--type experiment. However, with present limits on $\theta\leq (10\mbox{
TeV})^{-2}$ (obtained from studies of interaction with matter~\cite{r12}), visible light
(wavelength~$\sim10^{-5}$~cm) and attainable magnetic fields ($\sim1$~tesla) would lead to
the shift of 1~fringe on an optical length of 1~parsec ($10^{18}$~cm). Obviously more advanced
technology is required to render feasible this test of noncommutativity! 

Note that both polarizations undergo the same velocity change. This is in contrast to a
previous model for violating Lorentz symmetry in electrodynamics, wherein a spatial
Chern-Simons was added to the Maxwell action~\cite{r13}. In this modification the theory
remains linear; eqs.~\refeq{e33n} involve $\vec E$ and $\vec B$ [i.e., the constituent relations
\refeq{e34n} take on their $\theta=0$ value] but the right side of \refeq{e33a} acquires a term
proportional to~$\vec B$. The net result is that plane waves with different polarizations travel
with different velocities, producing a Faraday-type rotation, which can be ruled out
experimentally~\cite{r13,r14}. 

Although a Faraday rotation does not occur in the noncommutative  generalization of
Maxwell theory,
\refeq{e31n}, the following  is noteworthy. Observe that the modification in
\refeq{e31n} involves two numbers (1/8 and 1/2). 	Since $\theta$ may be rescaled, only
their ratio \refeq{n4} matters. One can check that when the ratio departs from this particular 
value,  plane waves still solve the equations, but now the two polarizations travel at
different velocities. So noncommutativity is the unique modification to Maxwell theory [within
the class
\refeq{e31n}]  that avoids a Faraday rotation. 

A final comment: It  has been observed that the action \refeq{e31n} may be viewed as a
Maxwell action in a background gravitational field, which itself is constructed
from~$F_{\mu\nu}$. Moreover, the equations \refeq{e32n}--\refeq{e34n} coincide with the
geodetic equation in that gravitational field~\cite{r15}. This hints at a view that
noncommutativity reflects the presence of a ``medium'', and this theme will reappear in an
explicit realization of the Seiberg-Witten map within fluid mechanics.

\bigskip \noindent {\large\bfseries  V}\enspace 
 The next investigation I shall describe results in an explicit formula for
the Seiberg-Witten map. The derivation again makes use of the fluid analogy, but now we need
the second, alternative formulation of fluid mechanics, the so-called Euler formulation. This is
not a comoving description (like the Lagrange formulation), rather the experimenter observes
the fluid density $\rho$ and velocity $\vec v$ at a given point in space-time $(t,\vec r)$. 

In preparation for our derivation, I shall briefly outline the Euler point of view. Moreover,
later I shall rely on that outline to give a generalization of Eulerian fluid mechanics to the case
when a non-Abelian symmetry group acts on the various degrees of freedom (components) of
the fluid. 

A point particle in space-time is described by its space-time coordinat $X^\mu (\tau)$, which
depends on an evolution variable~$\tau$ that parameterizes the path. One expects that
$X^\mu$ satisfies some dynamical equation, which determines $\ddot X^\mu$,  but rather
than concentrating on that equation (this would  correspond to a Lagrange description) we
focus on the density and current associated with that particle. These are functions of $x^\mu
\equiv (t,\vec r)$ and are related to $X^\mu$ by the formulas
\begin{subequations}\label{e30nn}
\begin{align}
\rho(t,\vec r) &= \int \rd \tau \dot X^0(\tau)\, \delta^4\bigl( X^\mu(\tau) -
x^\mu\bigr)\label{e30nna}\\
 &= \int \rd \tau \dot X^0(\tau)\, \delta\bigl( X^0(\tau) - t\bigr)\, \delta^3 \bigl(
\vec X(\tau) - \vec r \bigr) \nonumber\\
\vec j (t,\vec r) &= \int \rd \tau \dot {\vec X} (\tau)\, \delta^4 \bigl( X^\mu(\tau) -
x^\mu\bigr) \label{e30nnb}\\
 &= \int \rd \tau \dot {\vec X} (\tau)\, \delta\bigl( X^0(\tau) - t\bigr)\, \delta^3 \bigl(
\vec X(\tau) - \vec r \bigr)\nonumber
\end{align}
\end{subequations}
We can choose the parameterization so that $X^0(\tau) = \tau$, whereupon \refeq{e30nn}
becomes
\begin{subequations}\label{e31nn}
\begin{align}
\rho(t,\vec r) &=  \delta^3 \bigl(\vec X(t) - \vec r \bigr) \label{e31nna}\\
\vec j (t,\vec r) &=  \dot {\vec X} (t)\,  \delta^3 \bigl(\vec X(t) - \vec r \bigr)\ .
\label{e31nnb}
\end{align}
\end{subequations}
Observe that either from the definition \refeq{e30nn} or from \refeq{e31nn} a continuity
equation for $(\rho, \vec j)$ follows identically:
\numeq{e32nn}
{
\frac\partial{\partial t} \rho (t,\vec r)+ \grad \cdot \vec j(t,\vec r)=0\ .
}
Note also that \refeq{e31nnb} defines the velocity~$\vec v$ as a function of $(t,\vec r)$: it is 
$\dot{\vec X}(t)$ with $\vec X(t)$ replaced by $\vec r$: $\vec j=\rho\vec v$. 

A second equation is obtained by differentiating  \refeq{e31nnb} with respect to time: 
\begin{subequations}\label{e33nn}
\begin{align}
\frac\partial{\partial t} \vec j  &= \rho\dot{\vec v} + \dot\rho  \vec v  = \rho \dot{\vec v} - \vec
v\grad
\cdot (\rho\vec v)\nonumber\\
 &= \dot{\vec X} \dot X^j \frac\partial{\partial X^j} \delta^3\bigl(\vec X(t)-\vec
r\bigr)
 + \mbox{force}\ . 
\label{e33nna}
\end{align}
Here ``force'' denotes a formula for $\ddot{\vec X}$, which specifies the dynamics that govern
particle motion. The next to last term may be rewritten as
$$
-\dot{\vec X} \dot X^j \frac\partial{\partial r^j} \delta^3\bigl(\vec X(t)-\vec r\bigr)
 = -\frac\partial{\partial r^j}  (\vec v v^j\rho)\ . 
$$
Thus putting everything together we find
\numeq{e33nnb}{
 \frac\partial{\partial t} \vec v + \vec v \cdot \grad \vec v = \frac1\rho \ \mbox{force.}
}
\end{subequations}

Eqs.~\refeq{e32nn} and \refeq{e33nnb} are the Eulerian  equations for a ``fluid'' composed of a
single particle!  When there are several particles, labeled by index~$n$, formulas
\refeq{e31nn} are replaced by 
\begin{subequations}\label{e34nn}
\begin{align}
\rho(t,\vec r) &= \sum_n \delta^3 \bigl(\vec X_n(t) - \vec r \bigr) \label{e34nna}\\
\vec j (t,\vec r) &=  \rho(t,\vec r) \vec v(t,\vec r)\nonumber\\
&= \sum_n  \dot {\vec X}_n (t)\,  \delta^3 \bigl(\vec X_n(t) - \vec r
\bigr) 
\label{e34nnb}
\end{align}
\end{subequations}
but the subsequent development is as before, culminating in \refeq{e32nn} and
\refeq{e33nnb}. Finally, for a fluid, the discrete label becomes a continuous parameter
specifying the fluid
$n\to \vec x$, $\vec X_n (t) \to \vec X(t,\vec x)$; the density and velocity expressions are the
continuum generalizations of \refeq{e34nn}
\begin{subequations}\label{e35nn}
\begin{align}
\rho(t,\vec r) &= \int \rd{^3x} \delta^3 \bigl(\vec X(t,\vec x) - \vec r \bigr) \label{e35nna}\\
\vec j (t,\vec r) &=  \rho(t,\vec r) \vec v(t,\vec r)\nonumber\\
&= \int \rd{^3x}  \dot {\vec X} (t,\vec x)\,  \delta^3 \bigl(\vec X(t,\vec x) - \vec r
\bigr) 
\label{e35nnb}
\end{align}
\end{subequations}
and once again the continuity \refeq{e32nn} and Euler \refeq{e33nnb} equations are
established. 

It is useful to understand in detail the content of \refeq{e35nn}, which expresses the relation
between the Lagrange variables [$\vec X(t)$]  and the Euler variables [$\rho(t,\vec r), \vec
v(t,\vec r)$]. Evidently the integration over the parameter~$\vec x$ evaluates the delta
function at a value of $\vec x = \vec\chi (t,\vec r)$ such that 
\numeq{e36nn}{
\vec X(t,\vec \chi) = \vec r
}
and also there is a Jacobian. Thus \refeq{e35nna} also states that
\numeq{e37nn}{
\frac1{\rho(t,\vec r)} = \det \frac{\partial X^i (t, \vec x)}{\partial x^j}\Bigr|_{\vec x = \vec\chi}
}
while \refeq{e35nnb} implies
\numeq{e38nn}{
\vec v(t,\vec r) = \dot X^i (t,\vec x) \Bigr|_{\vec x = \vec\chi}\ . 
}
Effectively the passage from Lagrange to Euler descriptions involves interchanging the
dependent variable $\vec X$ with the independent parameter~$\vec x$, which is
renamed~$\vec r$ with the help of $\vec\chi(t,\vec r)$. 

Finally we note that the theory is completed by giving a model for the force. This will include
the (negative) gradient of the pressure, frequently taken as a function of~$\rho$. If the fluid is
charged, we are dealing with megnetohydrodynamics: $1/\rho$ force includes the Lorentz
force: $\vec E + \vec v\times \vec B$, and $j^\mu = (\rho, \rho\vec v)$ is the source for the
electromagnetic fields in Maxwell's equations
\numeq{e40nn}{
\partial_\mu F^{\mu\nu} = j^\nu \ . 
}
Later I shall describe how this is generalized when an internal symmetry is present, and a
group index~$a$ labels various physical quantities. 

Both a configuration space Lagrangian/action formulation and a canonical Hamiltonian
formulation for equations \refeq{e32nn}, \refeq{e33nnb} with pressure and Lorentz forces,
and \refeq{e40nn} can be given \cite{r5}.

\bigskip \noindent {\large\bfseries  VI}\enspace 
 Let me now turn to the task of obtaining  an explicit formula for the
Seiberg-Witten map from the fluid analogy.  Actually, I shall present the inverse map,
expressing commuting fields in terms of noncommuting ones; also the derivation will be
restricted to the case of two spatial dimensions [where $\theta^{ij}$ involves a single quantity
and there are \emph{no} integrability conditions on the differential
equation~\refeq{e30n}]. Higher dimensional cases have been been treated as well; the
development is more complicated. A few remarks about them will be given later; for a complete
discussion I refer you to the published literature~\cite{r4}.

The (inverse) Seiberg-Witten map, for the case of two spatial dimensions, can be extracted
from \refeq{e35nn}, which we rewrite as 
\begin{align}
j^\mu(t,\vec r) &= \int \rd {^2x}
\frac\partial{\partial t} X^\mu(t,\vec x) \delta^2 \bigl(\vec X(t,\vec x) - \vec r\bigr)
\nonumber\\
&[\dot X^0=1]\ .\label{e41n}
\end{align}
 Observe that the right side of
\refeq{e41n} depends on
$\hat {\vA}$ through
$\vec X$ [see
\refeq{e7}]. It is easy to check that the integral \refeq{e41n} is invariant under the
transformations \refeq{e4}; equivalently viewed as a function of $\hat {\vA}$, it is gauge
invariant   [see \refeq{e8}]. Owing to the conservation of $j^\mu, \partial_\mu j^\mu = 0$   [see
\refeq{e32nn}], its dual 
$\eps_{\alpha\beta\mu} j^\mu$ satisfies a conventional, commuting Bianchi identity, and
therefore can be written as the curl of an Abelian vector potential~$A_\alpha$, apart from
proportionality  and additive constants: 
\numeq{e42n}{
\begin{aligned}
\partial_\alpha A_\beta - \partial _\beta A_\alpha &{}+ \text{constant}\\ 
 \propto \eps_{\alpha\beta\mu} &{}\int \rd {^2x}
\frac\partial{\partial t} X^\mu
\delta^2(\vec X-
\vec r)\\
\partial_i A_j - \partial _j A_i &{}+ \text{constant} \\
 \propto \eps_{ij} &{}\int \rd {^2x}   \delta^2(\vec X- \vec r)
= \eps_{ij}
\rho\ .
\end{aligned}
 }
This is the (inverse) Seiberg-Witten map, relating the~$\vec A$ to~$\hat {\vA}$. 

Thus far the operator properties and the noncommutativity have not been taken into account.
To do so, first we reinterpret the integral over $\vec x$ as a trace over the Hilbert space that
realizes the noncommutativity of the coordinates (this is a standard procedure); second
we must provide an ordering for the $\delta$-function depending on the operator
$X^i = x^i +
\theta^{ij}
\hat A_j$.  This we do with the Weyl prescription by Fourier transforming. The  final operator
version of equation~\refeq{e42n}, restricted to the two-dimensional spatial components, reads
\numeq{e43n}{
\begin{gathered}
\int \rd {^2r}  e^{i\vec k\cdot\vec r} (\partial_i a_j - \partial_j a_i)
\qquad\qquad\qquad\\
\qquad= 
-\eps^{ij} \Bigl[ \int \rd {^2x}   e^{i\vec k\cdot\vec X} - (2\pi)^2\delta(\vec k)   \Bigr] \ .
\end{gathered}
}
Here the additive and proportionality constants are determined by requiring
agreement for weak noncommuting fields.

Formula~\refeq{e43n} has previously appeared  in a direct analysis of the Seiberg-Witten
relation~\cite{r16}. Now we recognize it as the (quantized) expression relating Lagrange and
Euler formulations for fluid mechanics~\cite{r4}. 

In higher-dimensional cases, greater than~2,  there arises the need to define a generalized
current tensor, whose dual gives rise to a two-index tensor satisfying a Bianchi identity
(closed and exact 2-form). There is some ambiguity in the construction when quantum
mechanical ordering needs to be implemented. The ambiguity precisely parallels the ambiguity
in ``integrating'' the path-dependent Seiberg-Witten differential equation; see
Ref.~\cite{r4}. 

\bigskip \noindent {\large\bfseries  VII}\enspace 
 A natural question that comes at this stage is whether one can extend
the fluids--noncommuting fields analogy to include a non-Abelian symmetry group. This has
not been achieved thus far, but for a first step we need to construct non-Abelian fluid
mechanics. This has been done. I shall now present the theory, which arises very naturally
from non-Abelian point-particle mechanics, just in the same was as I presented earlier a
derivation of  an (Eulerian) fluid from point-particle mechanics~\cite{r17}. 

Clearly, a non-Abelian fluid will possess a current $J_a^\mu$ that carries an internal
symmetry index~$a$. In the presence of dynamical gauge fields (non-Abelian
magnetohydrodynamics) $J_a^\mu$ serves as source for these fields in a generalization
of~\refeq{e40nn},
\numeq{e44n}{
(D_\mu F^{\mu\nu})_a = J^\nu_a
}
and is covariantly conserved, for consistency with \refeq{e44n}
\numeq{e45n}{
(D_\mu J^\mu)_a = 0\ . 
}
Here $D_\mu$ is the covariant derivative  $(D_\mu)^{ac} = \partial_\mu \delta^{ac} + f^{abc}
A_\mu^b$, where the non-Abelian gauge potential $A_\mu^a$ is contracted with the structure
constants $f^{abc}$ of the non-Abelian group. The challenge for us is to discover how the
non-Abelian current determines a charge density and velocity, and what kind of Euler
equations are satisfied by these quantities. 

To discover this, we recall the dynamics of a non-Abelian point particle, as described by
Wong~\cite{r18}. The particle carries a coordinate $\vec X(t)$ and a
non-Abelian charge
$q_a(t)$. The charge and current densities read [compare \refeq{e31nn}]
\numeq{e46n}{
\begin{aligned}
\rho_a(t,\vec r) &= q_a (t) \delta^3\bigl(\vec X(t) -\vec r\bigr)\\
\vec j_a(t,\vec r) &= q_a (t)\dot{\vec X}(t) \delta^3\bigl(\vec X(t) -\vec r\bigr)\ .
\end{aligned}
}
Covariant conservation is assured because the charge $q_a(t)$ is assumed to obey the Wong
equation 
\begin{align}
&\dot q_a(t) + f^{abc} \dot X^\mu (t) A_\mu^b \bigl(t,\vec X(t)\bigr) q^c (t) = 0
\nonumber\\
&\qquad[\dot X^0=1]\ .\label{e47n}
\end{align}
Moreover, particle acceleration is determined by the equation satisfied by $\vec X(t)$, e.g.,
\numeq{e48n}{
\dot P_\mu = q_a F_{\mu\nu}^a \dot X^\nu 
}
where $P_\mu$ is the particle 4-momentum. A Lagrangian/action and a Hamiltonian canonical 
formulation of these equations can be given; it involves the Kirilov-Kostant 1-form
on the Lie algebra~\cite{r19n}.

The passage to a fluid is now clear.  First we consider a finite number of particles, and introduce
a particle label~$n$: $X_n^\mu, P_\mu^n, q_a^n$. Next we pass to the continuum $n\to\vec x$:
$X^\mu(t,\vec x), P_\mu(t,\vec x), q_a (t,\vec x)$ -- these are the Lagrange variables. The
non-Abelian charge and current densities are defined as in~\refeq{e35nn}
\begin{subequations}\label{e49n}
\begin{align}
\rho_a(t,\vec r) &= \int \rd{^3x} q_a (t,\vec x) \delta^3\bigl(\vec X(t,\vec x) - \vec r\bigr)
\label{e49na}\\
\vec j_a(t,\vec r)&=\label{e49nb} \\ 
& \int \rd{^3x} q_a (t,\vec x) \dot{\vec X}(t,\vec x)
\delta^3\bigl(\vec X(t,\vec x) -\vec r\bigr)\ .
\nonumber
\end{align}
\end{subequations}
This remains covariantly conserved provided the Lagrangian non-Abelian charge variable
$q_a (t,\vec x)$ satisfies
\begin{align}
&\frac\partial{\partial t} q_a(t, \vec x)\label{e50n}\\
 &\qquad{}+ f_{abc} \dot X^\mu (t,\vec x)
A^b_\mu
\bigl(t, \vec X(t,\vec x)\bigr) q^c (t,\vec x) = 0\ . \nonumber
\end{align}
 
The important feature of these formulas is that by virtue of the $\delta$-function they
factorize into an Abelian current and a non-Abelian charge factor
\begin{subequations}\label{e51n}
\begin{align}
\rho_a(t,\vec r) &=   Q_a (t,\vec r) \rho(t,\vec r)
\label{e52na}\\
\vec j_a(t,\vec r) &= Q_a (t,\vec r) \vec j(t,\vec r)\nonumber\\
&= Q_a (t,\vec r) \rho(t,\vec r)  \vec v(t,\vec r) 
\label{e51nb} 
\end{align}
\end{subequations}
with $Q_a$ being the Eulerian charge, related to the Lagrange charge by
\numeq{e52n}{
Q_a (t,\vec r)  = q_a  (t,\vec x) \Bigr|_{\vec x = \vec \chi(t,\vec r)}
}
[see \refeq{e36nn}--\refeq{e38nn}]. Moreover, the Abelian current is still given by
\refeq{e35nn}, so it satisfies the Abelian continuity equation, and $\vec v$ satisfies an Euler
equation that reflects the underlying particle dynamics. Finally, the Eulerian charge
\refeq{e51n}  obeys the fluid Wong equation that follows from \refeq{e50n} and \refeq{e52n}
or alternatively from \refeq{e51n} when it is remembered that $j_a^\mu = Q_a j^\mu$ and
$j^\mu$ are respectively conserved covariantly and ordinarily:
\numeq{e53n}{
\frac\partial{\partial t}  Q_a + \vec v\cdot \grad Q_a =  f_{abc} Q^b (A_0^c - \vec v\cdot \vec
A^c)\ .  }
Non-Abelian magnetohydrodynamics then is described by the field equations \refeq{e44n},
\refeq{e45n}, by  the factorization \refeq{e51n}, by the fluid Wong equation \refeq{e53n},
and by the Euler equation
\numeq{e54n}{
\frac\partial{\partial t} \vec v + \vec v\cdot \grad \vec v = q^a \vec E_a + \vec v \times q^a
\vec B_a +
\frac1\rho \mbox{force} }
where we have explicitly exhibited the contribution to the force term of the non-Abelian
Lorentz force, involving the non-Abelian electric $\vec E_a$ and magnetic $\vec B_a$ fields. 
A Lagrangian for these equations makes use of
a field-theoretic generalization of the Kirilov-Kostant 1-form~\cite{r17}.

In the above approach a single density and velocity describe all the group components, i.e.,
$\rho$ and $\vec v$ are ``$a$''-independent. It is possible to give a more elaborate treatment
where different $\rho$ and $\vec v$ are associated with different elements of the Cartan
subalgebra of the group. Thus for SU(2), with one Cartan element, there would still be only a
single density and velocity describing all three group degrees of freedom.  On the other hand,
for SU(3) the more refined treatment uses two distinct densities and velocities. For this
elaboration I refer you to the literature \cite{r17}. 

Our approach to fluid was motivated by an underlying particle picture, as in the passage from
\refeq{e31nn} to \refeq{e35nn}. There is also a field-based approach, which is illustrated by
the fluid interpretation of the Schr\"odinger equation, due to Madelung~\cite{r19}. Observe
that the Schr\"odinger equation for a particle in an electromagnetic field derived from  
potentials
$(\phi,
\vec A)$ reads
\numeq{e55n}{
i\hbar \frac\partial{\partial t} \Psi = \fract12 \Bigl(\frac\hbar i \grad - e\vec A\Bigr)^2 \Psi
 + e\phi\Psi\ . 
}
If we define $\Psi = \rho^{1/2} e^{i\theta/\hbar}$, then the real and imaginary parts of
\refeq{e55n} become 
\begin{subequations}\label{e56n}
\begin{gather}
\frac\partial{\partial t} \rho + \grad\cdot \bigl((\grad\theta - e\vec A)\rho\bigr) = 0
\label{e56na}\\
\frac\partial{\partial t} \theta + \fract12 (\grad\theta - e\vec A)^2 = -e\phi + \hbar^2 M
\label{e56nb}\\
M\equiv \fract14 \frac{\grad\rho}\rho - \fract18 \frac{(\grad\rho)^2}{\rho^2}\ .
\end{gather}
\end{subequations}
The first of these suggests identifying $\grad\theta-e\vec A$ as the velocity, so that the
continuity equation \refeq{e32nn} is regained. Then upon taking the gradient of the second
equation, we arrive at
\numeq{e57n}{
\frac\partial{\partial t} \vec v + \vec v\cdot \grad \vec v = e(\vec E + \vec v\times\vec B) +
\frac1\rho \mbox{force}
 }
where ``force'' is the so-called Madelung $O(\hbar^2)$ quantum force given by $\rho\grad M$. 
Thus we see that Eulerian fluid mechanics emerges within this reanalysis of Schr\"odinger
theory. 

One may therefore assay a field-based non-Abelian fluid mechanics by beginning with a
Schr\"odinger theory for a  multicomponent wave function transforming under some
non-Abelian group. What emerges is a fluid mechanics that is much more complicated than
the particle-based model that we presented above. The crucial difference is that here there is
no simple factorization of the current, nor does a Wong equation hold~\cite{r17,r20}. 

It is unclear at present which (if any) of these non-Abelian generalizations for fluid mechanics
will prove to be the most useful. 
\vspace*{-.75pc}

\def\Journal#1#2#3#4{{\em #1} {\bf #2}, #3 (#4)}
\def\add#1#2#3{{\bf #1}, #2 (#3)}
\def\Book#1#2#3#4{{\em #1}  (#2, #3 #4)}
\def\Bookeds#1#2#3#4#5{{\em #1}, #2  (#3, #4 #5)}

\end{document}